\documentclass[letter]{aa}
\usepackage{txfonts}
\usepackage{graphicx}
\usepackage{subfigure}
\usepackage{natbib}
\bibpunct{(}{)}{;}{a}{}{,}

\def\chandra{{\it Chandra~}}

\def\swift{{\it Swift~}}

\def\xmm{{XMM-Newton~}}
\def\xmmk{{XMM-Newton}}

\def\m31{{M~31}}
\def\nova{{M31N~2007-11a~}}
\def\novak{{M31N~2007-11a}}
\def\source{{CXOM31 J004237.3+411710~}}
\def\sourcek{{CXOM31 J004237.3+411710}}
\def\msun{{$M_{\sun}$}}

\newcommand{\nh}{\hbox{$N_{\rm H}$}~}
\newcommand{\hcm}[1]{$\times 10^{#1}$ cm$^{-2}$}

\newcommand{\tpower}[1]{$\times 10^{#1}$}
\newcommand{\power}[1]{$10^{#1}$}

\begin{document}

\title{The very short supersoft X-ray state of the classical nova \nova\thanks{Partly 
   based on observations with \xmmk, an ESA Science Mission with instruments and contributions directly funded by ESA Member States and NASA}}

\author{M.~Henze\inst{1}
	\and W.~Pietsch\inst{1}
	\and G.~Sala\inst{2}
	\and M.~Della Valle\inst{3,4,5}
	\and M.~Hernanz\inst{6}
	\and J.~Greiner\inst{1}
	\and V.~Burwitz\inst{1}
	\and M.~J.~Freyberg\inst{1}
	\and F.~Haberl\inst{1}
	\and D.~H.~Hartmann\inst{7}
	\and P.~Milne\inst{8}
	\and G.~G.~Williams\inst{8}
}

\institute{Max-Planck-Institut f\"ur extraterrestrische Physik, 
	D-85748 Garching, Germany\\
	email: mhenze@mpe.mpg.de
	\and Departament de F\'isica i Enginyeria Nuclear, EUETIB (UPC/IEEC), Comte d'Urgell 187, 08036 Barcelona, Spain
	\and European Southern Observatory (ESO), D-85748 Garching, Germany
	\and INAF-Napoli, Osservatorio Astronomico di Capodimonte, Salita Moiariello 16, I-80131 Napoli, Italy
	\and International Centre for Relativistic Astrophysics, Piazzale della Repubblica 2, I-65122 Pescara, Italy
	\and Institut de Ci\`encies de l'Espai (CSIC-IEEC), Campus UAB, Fac. Ci\`encies, E-08193 Bellaterra, Spain
	\and Department of Physics and Astronomy, Clemson University, Clemson, SC 29634-0978, USA
	\and Steward Observatory, 933 North Cherry Avenue, Tucson, AZ 85721, USA
}

\date{Received 24 December 2008 / Accepted 13 March 2009}

\abstract
{Short supersoft X-ray source (SSS) states (durations $\le $ 100 days) of classical novae 
(CNe) indicate massive white dwarfs that are candidate progenitors of supernovae type Ia.}
{We carry out a dedicated optical and X-ray monitoring program of CNe in the bulge of \m31.}
{We discovered \nova and determined its optical and X-ray light curve. We used the robotic 
Super-LOTIS telescope to obtain the optical data and \xmm and \chandra observations to discover 
an X-ray counterpart to that nova.}
{Nova \nova is a very fast CN, exhibiting a very short SSS state with an appearance time of 6--16 days after outburst and a turn-off time of 45--58 days after outburst.}
{The optical and X-ray light curves of \nova suggest a binary containing a white dwarf with $M_{WD} >$ 1.0 \msun.}


\titlerunning{Short SSS state of \nova}

\maketitle

%
%
\section{Introduction}
%
Classical novae (CNe) are thermonuclear explosions on the surface of white 
dwarfs (WDs) in cataclysmic binaries that result from the transfer of matter 
from the companion star to the WD. The transferred hydrogen-rich matter accumulates 
on the surface of the WD until hydrogen ignition starts a thermonuclear runaway 
in the degenerate matter of the WD envelope. The resulting expansion of the hot 
envelope causes the brightness of the WD to rise by a typical outburst amplitude of $\sim$ 12 magnitudes 
within a few days, and mass to be ejected at high velocities \citep[see][and references 
therein]{2005ASPC..330..265H,1995cvs..book.....W}. However, a fraction of the hot 
envelope can remain in steady hydrogen burning on the surface of the WD 
\citep{1974ApJS...28..247S,2005A&A...439.1061S}, powering a supersoft X-ray source 
(SSS) that can be observed directly once the ejected envelope becomes sufficiently 
transparent \citep{1989clno.conf...39S,2002AIPC..637..345K}. In this paper, we define the term ``appearance of the SSS'' from an observational point of view as the time when the SSS becomes visible to us, due to the decreasing opacity of the ejected material. This is not the same as the actual turn-on of the H-burning on the WD surface at the beginning of the thermonuclear runaway. The turn off of the SSS, on the other hand, does not suffer from extinction effects and clearly marks the actual turn off of the H-burning in the WD atmosphere and the disappearance of the SSS.

The duration of the supersoft phase is related to the amount of H-rich matter that is 
{\it not} ejected and on the luminosity of the white dwarf. More massive WDs need to accrete less matter to initiate the 
thermonuclear runaway, because of their higher surface gravity \citep{1998ApJ...494..680J}. 
In general, more massive WDs retain less accreted mass after the explosion, although this also 
depends on the accretion rate, and reach higher luminosities \citep{2005ApJ...623..398Y}. Thus, the 
duration of the SSS state is inversely related to the mass of the WD. In turn, the transparency 
requirement mentioned above implies that the appearance time is determined by the fraction of mass 
ejected in the outburst \citep[see][]{2005A&A...439.1061S,1998ApJ...503..381T,2006ApJS..167...59H}. 
X-ray light curves therefore provide important clues to the physics of the nova outburst, 
addressing the key question of whether a WD accumulates matter over time to become a potential
progenitor for a type Ia supernova (SN-Ia). The duration of the SSS state provides the only 
direct indicator of the post-outburst envelope mass in CNe. For massive WDs, the expected SSS 
phase is very brief ($<$ 100 d) \citep{2005A&A...439.1061S,1998ApJ...503..381T} and could have 
easily been missed in previous surveys \citep[e.g.][]{2001A&A...373...63S,2004ApJ...609..735W,2005A&A...434..483P}.

There are only a few novae with very short SSS states known. One of them is the well-studied galactic nova RS~Oph. RS~Oph is a recurrent nova (RN), which underwent its last outburst in 2006. X-ray observations documented a rapid decline in supersoft emission at $\sim 90$ days after the optical outburst and a duration of the SSS phase of $\sim 60$ days \citep{2006ATel..838....1O,2007ApJ...659L.153H}. From theoretical models, \citet{2007ApJ...659L.153H} estimated a WD mass of $(1.35\pm0.01)$\msun. RNe are classified by the observational fact that they have had more than one recorded outburst. These objects are very good candidates for SN-Ia progenitors, as RNe are believed to contain massive WDs. However, one of the problems with this RNe $-$ SN-Ia connection was their rare occurrence in optical surveys \citep{1996ApJ...473..240D}. But hypothetical previous outbursts may have been missed, and then these objects were just classified as CNe in optical surveys. Interestingly, a very short SSS stage of a CN may indicate a very massive WD, which is capable of frequent outbursts, and thus is a potential RN.

Another example for a nova with short SSS state is the CN Nova LMC 2000, for which \citet{2003A&A...405..703G}, based on non-detections of supersoft emission, suggest an SSS phase shorter than seven weeks. The work of \citet{2007A&A...465..375P} already discusses nova M31N~2004-11f as an object with a very short SSS state of 35--55 days. This nova is an RN.

This work presents the results of a new monitoring strategy tailored to CNe with short SSS 
states. We carried out combined \xmmk/\chandra monitoring of the central region of \m31 
\citep[distance 780 kpc,][]{1998AJ....115.1916H,1998ApJ...503L.131S} from November 2007 to 
mid-February 2008. Individual observations were separated by just 10 days. Section \ref{sec:obs} 
provides detailed information on the X-ray and optical data sets. Results are presented 
in Sect.\,\ref{sec:results} and discussed in Sect.\,\ref{sec:discuss}.

%
%
\section{Observations and data analysis}
\label{sec:obs}
%
\subsection{Optical data}
\label{sec:obs_opt}
The optical data used in this work were obtained with Super-LOTIS
\citep[Livermore Optical Transient Imaging System,][]{2008AIPC.1000..535W}, a 
robotic 60 cm telescope equipped with an E2V CCD (2kx2k) located at Steward Observatory, 
Kitt Peak, Arizona, USA. Starting in October 2007, the telescope was used every 
good night to monitor the bulge of \m31. Using four Super-LOTIS fields (field of view: 
$17\arcmin \times 17\arcmin$), we covered an area of $\sim 34\arcmin \times 34\arcmin$ 
centered on the core of M\,31. The pixel scale is 0.496\arcsec/pixel and the typical 
limiting magnitude is 19 mag, in Johnson R. The data were reduced by a semi-automatic routine and the astrometric and photometric calibration utilizes the 
\m31 part of the Local Group Survey \citep[LGS,][]{2006AJ....131.2478M}. Typical values 
of Super-LOTIS astrometric and photometric $1\sigma$ accuracies are $0\,\farcs25$ and 
0.25 mag, respectively, averaged over the whole magnitude range. 

\subsection{X-ray data}
\label{sec:obs_xray}
The X-ray data were taken as part of our \m31 nova monitoring 
project\footnote{http://www.mpe.mpg.de/$\sim$m31novae/xray/index.php} 
using the \chandra High-Resolution Camera Imaging Detector (HRC-I) and the \xmm European Photon Imaging Camera (EPIC) PN detector. The individual observations used here are summarized in Table\,\ref{table:xray}, which lists the telescopes and instruments used, the observation 
identifications (ObsIDs), the dates, source count rates, and luminosities or 
upper limits. With the help of the mission count rate simulator PIMMS v3.9a, we computed the 
energy conversion factors needed to transform count rates into unabsorbed fluxes for the 
individual telescopes.

For \xmm data we applied background screening and used the XMMSAS (\xmm Science Analysis Software) 
v8.0 tasks \texttt{eboxdetect} and \texttt{emldetect} to detect sources in the image and 
perform astrometry and photometry. For computing upper limits, we added an artificial detection 
at the position of the source to the \texttt{eboxdetect} list. 
This list was used as input for \texttt{emldetect} (with fixed positions and likelihood 
threshold of zero) to derive the observed flux, or an upper limit, for all objects in the list.

We reduced the \chandra observations with the CIAO v3.4 (Chandra Interactive Analysis of Observations) software package. The source detection was done with the CIAO tool \texttt{wavedetect}, and an adapted version of the XMMSAS tool \texttt{emldetect} was used to estimate background-corrected and exposure-corrected fluxes and count rates for the detected sources. We applied an astrometric calibration to the object list for each individual observation using the X-ray source catalog of \citet{2002ApJ...578..114K} to improve the astrometric accuracy of our measurements.

%
\begin{table*}[ht]
\caption{X-ray observations of \source.}
\label{table:xray}
\begin{center}
\begin{tabular}{lrrrrrr}\hline\hline \noalign{\smallskip}
	Telescope/Instrument & ObsID & Exp. Time$^a$ & Start Date & Offset$^b$ & Count Rate & L$_{0.2-1.0}$ $^c$\\
	 & & [ks] & [UT] & [d] & [ct s$^{-1}$] & [erg s$^{-1}$]\\ \hline \noalign{\smallskip}
	\chandra HRC-I & 8526 & 19.9 & 2007-11-07.64 & 5.89 & $<$ 8.7 \tpower{-4} & $<$ 4.0 \tpower{36}\\
	\chandra HRC-I & 8527 & 20.0 & 2007-11-17.76 & 16.01 & (17.4 $\pm$ 1.1) \tpower{-3} &  (8.1 $\pm$ 0.5) \tpower{37}\\
	\chandra HRC-I & 8528 & 20.0 & 2007-11-28.79 & 27.04 & (4.3 $\pm$ 0.5) \tpower{-3} &  (2.0 $\pm$ 0.2) \tpower{37} \\
	\chandra HRC-I & 8529 & 18.9 & 2007-12-07.57 & 35.82 & (4.7 $\pm$ 0.5) \tpower{-3} &  (2.2 $\pm$ 0.2) \tpower{37}\\
	\chandra HRC-I & 8530 & 19.9 & 2007-12-17.49 & 45.29 & (3.0 $\pm$ 0.4) \tpower{-3} &  (1.4 $\pm$ 0.2) \tpower{37}\\
	\xmm EPIC PN & 0505720201 & 22.2 & 2007-12-29.57 & 57.80 & $<$ 8.7 \tpower{-4} &  $<$ 2.7 \tpower{35}\\ 
	\xmm EPIC PN & 0505720301 & 21.9 & 2008-01-08.29 & 67.53 & $<$ 3.0 \tpower{-3} &  $<$ 5.4 \tpower{35}\\ 
	\xmm EPIC PN & 0505720401 & 18.1 & 2008-01-18.63 & 77.87 & $<$ 2.2 \tpower{-3} &  $<$ 6.0 \tpower{35}\\ 
	\xmm EPIC PN & 0505720501 & 17.2 & 2008-01-27.94 & 87.17 & $<$ 2.0 \tpower{-3} &  $<$ 1.3 \tpower{36}\\ 
	\xmm EPIC PN & 0505720601 & 17.3 & 2008-02-07.20 & 97.44 & $<$ 4.3 \tpower{-3} &  $<$ 4.2 \tpower{35}\\ 
\noalign{\smallskip} \hline
\end{tabular}
\end{center}
\noindent
Notes:\hspace{0.3cm} $^a $: Dead-time corrected\\
\hspace*{1.1cm} $^b $: Time in days after the outburst of nova \nova in the optical \citep{2007ATel.1257....1P} on 2007 Nov 2.28 UT\\
\hspace*{1.5cm} (MJD = 54406.2); uncertainties are 0.45 days, according to Sect.\,\ref{sec:results}\\
\hspace*{1.1cm} $^c $: X-ray luminosities and 3$\sigma$ upper limits (unabsorbed, 0.2 - 1.0 keV) estimated using a blackbody model with kT = 50eV\\
\hspace*{1.5cm} and \nh = 6.7\hcm{20} (see Sect.\,\ref{sec:obs_xray})\\
\end{table*}
%

\section{Results}
\label{sec:results}
\begin{figure}
	\resizebox{\hsize}{!}{\includegraphics[angle=90]{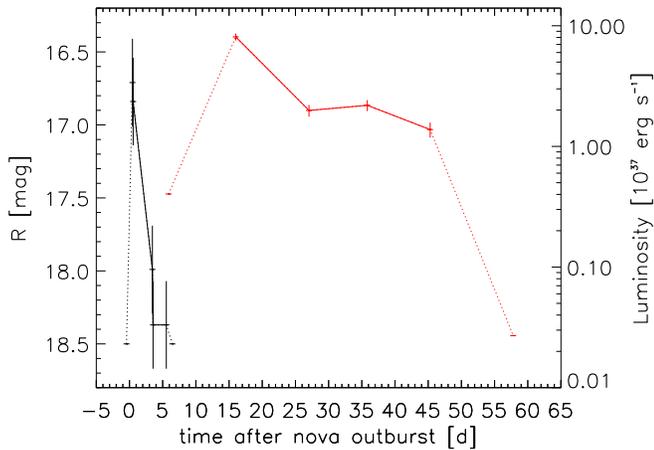}}
	\caption{Optical (black symbols and left y-axis) and X-ray (red symbols and right y-axis) light curve of \nova. Detections are marked by crosses, the sizes of which correspond to the errors in magnitude/luminosity and time after outburst (MJD = 54405.75). Upper limits are represented by horizontal bars that show the uncertainty in time after outburst. Detections are connected with solid lines, upper limits to detections by dotted lines.}
	\label{fig:lc}
\end{figure}

%
In our Super-LOTIS monitoring data of 2007 Nov 2.28 UT, we discovered \nova as a previously unknown nova candidate \citep{2007ATel.1257....1P}. The position of \nova is RA = 00h42m37.29s, Dec = +41$\degr$17$\arcmin 10\,\farcs3$ (J2000, accuracy of 0.2"), which is 1$\arcmin$ 19$\arcsec$ west and 1$\arcmin$ 2$\arcsec$ north of the core of M 31. The detected R band magnitudes are 16.7 (2007 Nov 2.28), 16.8 (2.37), 18.0 (5.27), 18.4 (5.35), 18.4 (7.35). The optical light curve of \nova is given in Fig.\,\ref{fig:lc}. We estimate the photometric accuracy of our measurements of \nova to be $\sim$ 0.3 mag, which is slightly less accurate than average since the object is located very near the center of \m31, and measurements suffer from strong background light. The outburst date of \nova is well-constrained, since it was first detected on MJD = 54406.2 but not on MJD = 54405.3. Therefore, we estimate MJD = 54405.75 for the actual outburst, with an error of $\pm0.45$ days. The horizontal error bars in Fig.\,\ref{fig:lc} represent this uncertainty for the days after the nova outburst.

In the second HRC-I monitoring observation of our \m31 monitoring campaign (starting 2007 
Nov 17.76), \source was detected as a relatively bright source ($348\pm23$ counts), whereas in the 
first HRC-I observation ten days earlier, no source was detected at this position. There is no 
known X-ray source within $25\arcsec$ of the position of \sourcek, in previous \m31 \xmm and 
\chandra X-ray source catalogs \citep{2002ApJ...578..114K,2002ApJ...577..738K,
2004ApJ...609..735W,2005A&A...434..483P,2008A&A...480..599S}. The luminosity of the source 
declined during the three subsequent \chandra observations (Nov 28.79, Dec 07.57, Dec 17.49). 
In the first (Dec 29) and all following \xmm monitoring observations, \source is no longer detected. 
The X-ray light curve of \source (Fig.\,\ref{fig:lc}) shows that this object is a very fast 
transient. Count rates and luminosities (see below) are given in Table\,\ref{table:xray}.

The \chandra position of \source is RA = 00h42m37.29s, Dec = +41$\degr$17$\arcmin 10\,\farcs2$ 
(J2000, $3 \sigma$ error of $0\,\farcs8$), which is in excellent agreement 
(distance of $0\,\farcs1$), within the errors, with the optical position of \nova. The X-ray 
position uncertainties are only statistical and do not include a possible systematic offset 
between the catalog of \citet{2002ApJ...578..114K} and the LGS \citep{2006AJ....131.2478M}.

In Fig.\,\ref{fig:lc} we give the X-ray light curve of \source with respect to the outburst of \nova. The delay times to the nova outburst given in Table\,\ref{table:xray} for upper limits and detections imply that the appearance of the 
SSS happened between 6 and 16 days after the optical outburst and that the SSS state lasted between 
29 and 52 days. 

There is no detailed spectral information about \source, because of the very limited energy 
resolution of the HRC detector and the source not being visible in our \xmm monitoring 
observations. Also, observations with the \swift X-ray telescope from 2007 Nov 24 to Dec 2 
(ObsIDs 031028001 to 031028008) and with the \chandra Advanced CCD Imaging Spectrometer ACIS-I 
on 2007 Nov 27 (ObsID 8187) were not deep enough to help with the spectral classification. (There 
is also no UV detection of \source with the \swift UV/Optical Telescope or the \xmm Optical 
Monitor.) We therefore used the hardness ratios from the HRC-I described in ``The Chandra Proposers' 
Observatory Guide"\footnote{see http://cxc.harvard.edu/proposer/POG/html/index.html\\ \hspace*{0.5cm} 
chapter 7.6} for ObsID 8527 derived from the rates in the bands S, M, and H (channels 1:100, 100:140, 
and 140:255, respectively). We derived ratios S/M = $-0.10\pm0.15$, M/H = $0.09\pm0.15$, indicating
a very soft spectrum. To compare luminosities from \chandra and \xmm upper limits, we assumed a blackbody model with $kT = 50$eV and absorption column of 6.7\hcm{20}, the galactic foreground 
\nh towards \m31. An absorbed blackbody model is the simplest way of describing an SSS spectrum \citep[see also][]{2007A&A...465..375P}. The temperature is an average value for the \m31 novae described in the works of \citet{2007A&A...465..375P,2005A&A...442..879P}. We assume a source distance of 780 kpc 
\citep{1998AJ....115.1916H,1998ApJ...503L.131S}. The usage of a generic spectral model, Although only done to compare the different instruments, using a generic spectral model may of course produce significant luminosity uncertainties. Therefore, in Table\,\ref{table:ecf} we give the HRC energy conversion factors (count rate / flux, computed using PIMMS), as well as unabsorbed HRC peak 
luminosities and PN upper limits of \source for different blackbody temperatures and an \nh fixed to the Galactic foreground value. The table shows that the source can be classified as a luminous SSS (L$_{0.2-1.0}$ $>$ \power{37} 
erg s$^{-1}$) independent of the assumed blackbody temperature.

%
\begin{table}[ht]
\caption{HRC energy conversion factors (ecf's) and unabsorbed HRC and PN luminosities for different blackbody temperatures.}
\label{table:ecf}
\begin{center}
\begin{tabular}{cccc}
	\hline\noalign{\smallskip}
	\hline\noalign{\smallskip}
	$kT$ & ecf$_{HRC}$ & $L_{0.2-1.0}^{HRC}$ $^a$ & $L_{0.2-1.0}^{PN}$ $^b$\\ \noalign{\smallskip}
	$ $ [eV] & [\power{10} ct cm$^2$ erg$^{-1}$] & [\power{37} erg s$^{-1}$] & [\power{35} erg s$^{-1}$]\\
	\noalign{\smallskip}\hline\noalign{\smallskip}
	40 & 1.16 & 10.9 $\pm$ 0.7 & 3.6\\
	50 & 1.57 & 8.1 $\pm$ 0.5 & 2.7\\
	60 & 2.09 & 6.1 $\pm$  0.4 & 2.3\\
	70 & 2.71 & 4.7 $\pm$ 0.3 & 2.1\\
	 \hline
\end{tabular}
\end{center}
\noindent
Notes:\hspace{0.3cm} $^a $: For ObsID 8527 (see Table\,\ref{table:xray})\\
\hspace*{1.1cm} $^b $: For ObsID 0511380201 (see Table\,\ref{table:xray})\\
\end{table}
%

\section{Discussion}
\label{sec:discuss}
%
From the positional and temporal agreement and from the HRC-I hardness ratios, we 
infer that \source and \nova are the same object.

\begin{figure}
	\resizebox{\hsize}{!}{\includegraphics[angle=0]{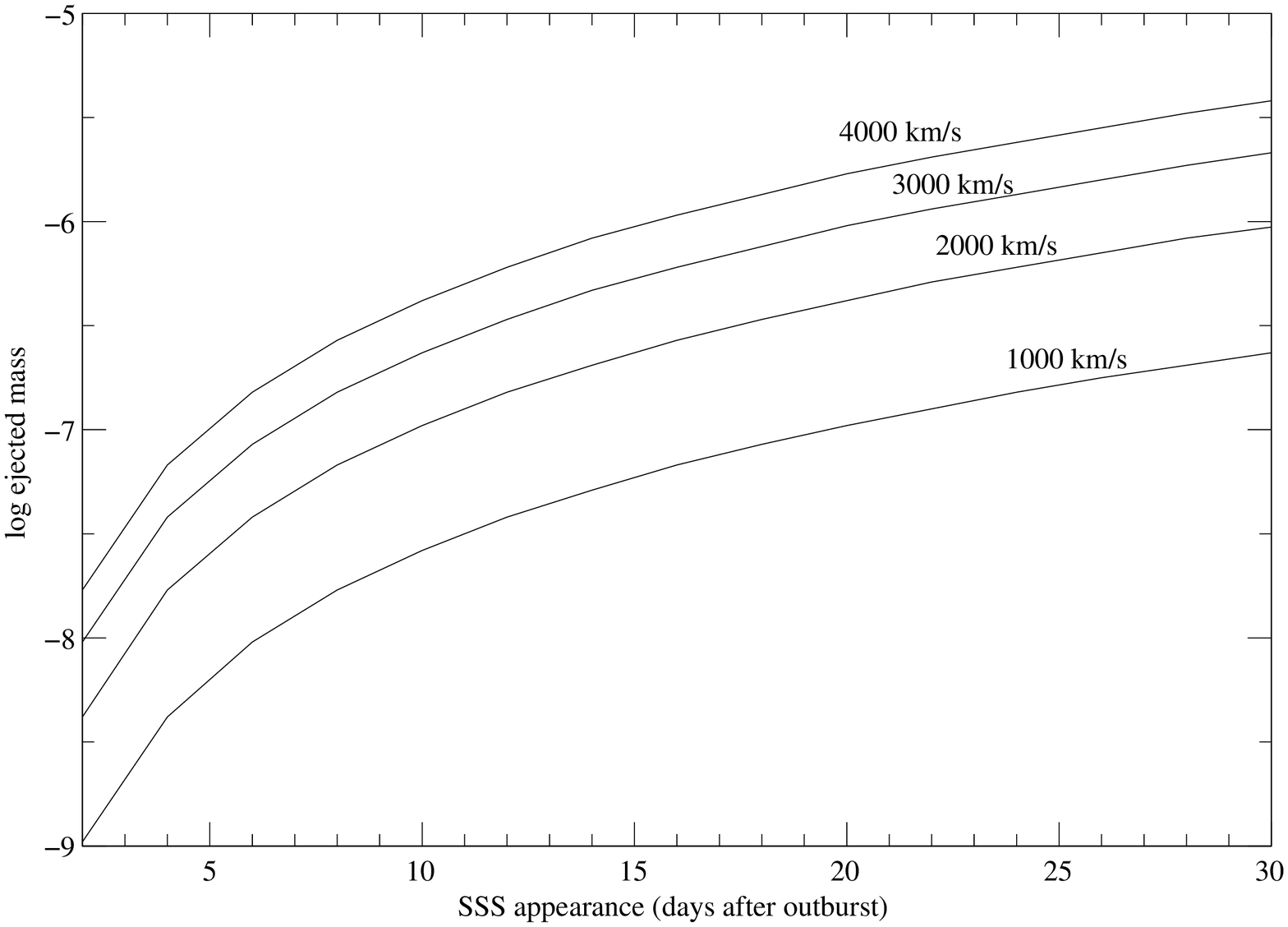}}
	\caption{Relation of ejected mass (in solar units) and SSS appearance time for different 
        expansion velocities $v$ (see Sect.\,\ref{sec:discuss}).}
	\label{fig:param}
\end{figure}

The ejected mass in the nova outburst can be estimated from the start date of the 
SSS phase and from the expansion velocity of the ejected material. The decrease in the optical thickness of the expanding ejecta is responsible 
for the rise in the X-ray light curve of the post-outburst novae, as shown for V1974 Cyg 
\citep{1996ApJ...463L..21S,1996ApJ...456..788K}. Assuming that the material ejected by the 
nova explosion forms a spherical, homogeneous shell expanding at constant velocity $v$, 
the hydrogen mass density of the shell will evolve in time $t$ as $\rho=M^{ej}_{H} / 
(\frac{4}{3}\pi v^{3}t^{3})$ where $M^{ej}_{H}$ is the ejected hydrogen mass 
\citep{1996ApJ...456..788K}. The column density of hydrogen evolves with time as
$N_{H} ({\rm cm}^{-2})= M^{ej}_{H} / (\frac{4}{3}\pi u v^{2}t^{2})$, where 
$u=1.673\times10^{-24}$ g is the atomic mass unit. We assume a typical value 
for the expansion velocity of 2000 km s$^{-1}$, since we do not have an optical spectrum 
for \nova, and that the SSS turns on when the absorbing hydrogen column density decreases 
to $\sim10^{21}$ cm$^{-2}$. With this, the appearance time of the SSS source (between 6 and 16 days) 
constrains the ejected mass to the range $(0.4-3)\times10^{-7}$\msun. This is about two orders 
of magnitude lower than typical ejected masses determined in a similar way for M31 novae 
\citep{2007A&A...465..375P}. However, the unknown expansion velocity error adds additional uncertainties to the estimate of $M^{ej}_{H}$. Figure\,\ref{fig:param} shows the range of $M^{ej}_{H}$ for 
different expansion velocities as a function of the SSS appearance time.

The turn-off time of the SSS phase allows us to estimate the amount of hydrogen-rich material 
burned on the WD surface, $M^{burn}=(L\Delta t) / (X_H \epsilon)$, where $L$ is the bolometric 
luminosity, $\Delta t$ the duration of the SSS phase, $X_H$ the hydrogen fraction of the 
burned material, and $\epsilon=5.98\times10^{18}$ erg g$^{-1}$. We compute the burned mass during 
the short SSS phase of \nova, assuming a bolometric luminosity of $3\times10^4L_{\odot}$ and a 
hydrogen mass fraction $X_H=0.5$. The SSS turn off at 45--58 days after the optical outburst 
constrains the burned mass to the range $(8-10)\times10^{-8}$\msun. It is noteworthy that this 
is within a factor 2--3 the same as the ejected mass. We used typical values for $L$ and $X_H$, since we do not know the actual values for \nova.

\nova has the shortest SSS phase observed so far, which can be caused either by a WD with $M_{WD} >$ 1.1 \msun for a standard hydrogen fraction in the envelope or by very hydrogen-poor envelope in a WD with $M_{WD} \sim$ 1.0 \msun. The mass of the stable H-burning envelope decreases with increasing WD mass, with massive WDs showing 
the shortest SSS phases. Massive WDs also eject less massive shells during outburst, leading to a faster 
rise in the SSS emission.

From the optical light curve we find a decline rate of $\sim 0.6$ mag/d. Extrapolating the 
magnitude to the day before the first observation (MJD = 54405.3) leads to an estimated maximum 
magnitude of $m_R \sim 16.1$, which implies an absolute magnitude at maximum of $M_R \sim -8.4 
\pm 0.4$ at the distance of \m31. Within the uncertainties, this value matches the maximum 
magnitude versus rate of decline (MMRD) relationship obtained by \citet{1995ApJ...452..704D}. 
Assuming that the strength of a nova outburst (for a given temperature of the WD and accretion 
rate) mostly depends on the mass of the WD \citep[see e.g.]
[and references therein]{1992ApJ...393..516L,1995ApJ...452..704D,2002AIPC..637..443D}, one can estimate WD masses from optical nova light curves. However, \citet{1992ApJ...393..516L} mentions that great caution should be exercised when applying his maximum magnitude vs. WD mass relation to individual novae, since there are weaker dependences of the nova luminosity on the magnetic field of the WD and the enrichment of the envelope by heavy elements, which are neglected in his formula. Using our optical light curve, we compute that\nova likely originated from a WD with $M_{WD} \sim$ 0.9--1.15 \msun. This result is consistent with the X-ray observations.

As mentioned before, many of the novae with known short SSS phases turned out to be RNe. To check for the possibility that \nova is an RN, we looked for historical novae with positions close to \novak. In the compiled on-line nova catalog of \citet{2007A&A...465..375P}\footnote{http://www.mpe.mpg.de/$\sim$m31novae/opt/m31/index.php} there are two historical novae within a radius of $\sim 20 \arcsec$ around \novak. \citet{2000AstL...26..433S} found a nova (M31N~1998-07n) at a distance of about $15 \arcsec$, and \citet{1929ApJ....69..103H} reported a nova (M31N~1920-10a) at a distance of about $20 \arcsec$ to \novak. The J2000 coordinates of M31N~1920-10a were taken from the Combined General 
Catalogue of Variable Stars \citep{2004yCat.2250....0S}. From the finding chart given by \citet{2000AstL...26..433S}, nova M31N~1998-07n and \nova do not seem to be identical. This agrees with the position accuracy of $0\,\farcs5$ given by \citet{2000AstL...26..433S}. For M31N~1920-10a no finding chart exists and the position given by \citet{1929ApJ....69..103H} has a minimum accuracy of $0\,\farcm1$. The actual unknown position accuracy of historical novae leaves the (small) possibility open that both novae in fact originated on the same WD. A relatively large position error for a historical nova could be possible, since e.g. the re-analysis of photographic plates by \citet{2008A&A...477...67H} gives several examples for historical novae with position errors of $\sim 20 \arcsec$ or more.
Clarification of this point could only come from re-examination of the original photographic 
plates used in the work of \citet{1929ApJ....69..103H}.

\begin{acknowledgements}
We wish to thank the anonymous referee for helpful comments that helped to improve the clarity of the paper. The X-ray work is based in part on observations with \xmmk, an ESA Science Mission with 
instruments and contributions directly funded by ESA Member States and NASA. The \xmm project 
is supported by the Bundesministerium f\"{u}r Wirtschaft und Technologie / Deutsches Zentrum 
f\"{u}r Luft- und Raumfahrt (BMWI/DLR FKZ 50 OX 0001), the Max-Planck Society, and the 
Heidenhain-Stiftung. M. Henze acknowledges support from the BMWI/DLR, FKZ 50 OR 0405. G.S. acknowledges support from grants AYA2008-04211-C02-01 and AYA2007-66256.
M. Hernanz acknowledges support from grants ESP2007-61593 and 2005-SGR00378. 
D.H. acknowledges internal funding from Clemson University for partial support of the 
operation of Super-LOTIS.
\end{acknowledgements}
\bibliographystyle{aa}

\begin{thebibliography}{36}
\expandafter\ifx\csname natexlab\endcsname\relax\def\natexlab#1{#1}\fi

\bibitem[{{Della Valle}(2002)}]{2002AIPC..637..443D}
{Della Valle}, M. 2002, in American Institute of Physics Conference Series,
  Vol. 637, Classical Nova Explosions, ed. M.~{Hernanz} \& J.~{Jos{\'e}},
  443--456

\bibitem[{{Della Valle} \& {Livio}(1995)}]{1995ApJ...452..704D}
{Della Valle}, M. \& {Livio}, M. 1995, \apj, 452, 704

\bibitem[{{Della Valle} \& {Livio}(1996)}]{1996ApJ...473..240D}
{Della Valle}, M. \& {Livio}, M. 1996, \apj, 473, 240

\bibitem[{{Greiner} {et~al.}(2003){Greiner}, {Orio}, \&
  {Schartel}}]{2003A&A...405..703G}
{Greiner}, J., {Orio}, M., \& {Schartel}, N. 2003, \aap, 405, 703

\bibitem[{{Hachisu} \& {Kato}(2006)}]{2006ApJS..167...59H}
{Hachisu}, I. \& {Kato}, M. 2006, \apjs, 167, 59

\bibitem[{{Hachisu} {et~al.}(2007){Hachisu}, {Kato}, \&
  {Luna}}]{2007ApJ...659L.153H}
{Hachisu}, I., {Kato}, M., \& {Luna}, G.~J.~M. 2007, \apjl, 659, L153

\bibitem[{{Henze} {et~al.}(2008){Henze}, {Meusinger}, \&
  {Pietsch}}]{2008A&A...477...67H}
{Henze}, M., {Meusinger}, H., \& {Pietsch}, W. 2008, \aap, 477, 67

\bibitem[{{Hernanz}(2005)}]{2005ASPC..330..265H}
{Hernanz}, M. 2005, in Astronomical Society of the Pacific Conference Series,
  Vol. 330, The Astrophysics of Cataclysmic Variables and Related Objects, ed.
  J.-M. {Hameury} \& J.-P. {Lasota}, 265

\bibitem[{{Holland}(1998)}]{1998AJ....115.1916H}
{Holland}, S. 1998, \aj, 115, 1916

\bibitem[{{Hubble}(1929)}]{1929ApJ....69..103H}
{Hubble}, E.~P. 1929, \apj, 69, 103

\bibitem[{{Jose} \& {Hernanz}(1998)}]{1998ApJ...494..680J}
{Jose}, J. \& {Hernanz}, M. 1998, \apj, 494, 680

\bibitem[{{Kaaret}(2002)}]{2002ApJ...578..114K}
{Kaaret}, P. 2002, \apj, 578, 114

\bibitem[{{Kong} {et~al.}(2002){Kong}, {Garcia}, {Primini}, {Murray}, {Di
  Stefano}, \& {McClintock}}]{2002ApJ...577..738K}
{Kong}, A.~K.~H., {Garcia}, M.~R., {Primini}, F.~A., {et~al.} 2002, \apj, 577,
  738

\bibitem[{{Krautter}(2002)}]{2002AIPC..637..345K}
{Krautter}, J. 2002, in American Institute of Physics Conference Series, Vol.
  637, Classical Nova Explosions, ed. M.~{Hernanz} \& J.~{Jos{\'e}}, 345

\bibitem[{{Krautter} {et~al.}(1996){Krautter}, {\"Ogelman}, {Starrfield},
  {Wichmann}, \& {Pfeffermann}}]{1996ApJ...456..788K}
{Krautter}, J., {\"Ogelman}, H., {Starrfield}, S., {Wichmann}, R., \&
  {Pfeffermann}, E. 1996, \apj, 456, 788

\bibitem[{{Livio}(1992)}]{1992ApJ...393..516L}
{Livio}, M. 1992, \apj, 393, 516

\bibitem[{{Massey} {et~al.}(2006){Massey}, {Olsen}, {Hodge}, {Strong},
  {Jacoby}, {Schlingman}, \& {Smith}}]{2006AJ....131.2478M}
{Massey}, P., {Olsen}, K.~A.~G., {Hodge}, P.~W., {et~al.} 2006, \aj, 131, 2478

\bibitem[{{Osborne} {et~al.}(2006){Osborne}, {Page}, {Beardmore}, {Goad},
  {Bode}, {O'Brien}, {Schwarz}, {Starrfield}, {Ness}, {Krautter}, {Drake},
  {Evans}, \& {Eyres}}]{2006ATel..838....1O}
{Osborne}, J., {Page}, K., {Beardmore}, A., {et~al.} 2006, The Astronomer's
  Telegram, 838, 1

\bibitem[{{Pietsch} {et~al.}(2007{\natexlab{a}}){Pietsch}, {Burwitz}, {Updike},
  {Milne}, {Williams}, \& {Hartmann}}]{2007ATel.1257....1P}
{Pietsch}, W., {Burwitz}, V., {Updike}, A., {et~al.} 2007{\natexlab{a}}, The
  Astronomer's Telegram, 1257, 1

\bibitem[{{Pietsch} {et~al.}(2005{\natexlab{a}}){Pietsch}, {Fliri}, {Freyberg},
  {Greiner}, {Haberl}, {Riffeser}, \& {Sala}}]{2005A&A...442..879P}
{Pietsch}, W., {Fliri}, J., {Freyberg}, M.~J., {et~al.} 2005{\natexlab{a}},
  \aap, 442, 879

\bibitem[{{Pietsch} {et~al.}(2005{\natexlab{b}}){Pietsch}, {Freyberg}, \&
  {Haberl}}]{2005A&A...434..483P}
{Pietsch}, W., {Freyberg}, M., \& {Haberl}, F. 2005{\natexlab{b}}, \aap, 434,
  483

\bibitem[{{Pietsch} {et~al.}(2007{\natexlab{b}}){Pietsch}, {Haberl}, {Sala},
  {Stiele}, {Hornoch}, {Riffeser}, {Fliri}, {Bender}, {B{\"u}hler}, {Burwitz},
  {Greiner}, \& {Seitz}}]{2007A&A...465..375P}
{Pietsch}, W., {Haberl}, F., {Sala}, G., {et~al.} 2007{\natexlab{b}}, \aap,
  465, 375

\bibitem[{{Sala} \& {Hernanz}(2005)}]{2005A&A...439.1061S}
{Sala}, G. \& {Hernanz}, M. 2005, \aap, 439, 1061

\bibitem[{{Samus} {et~al.}(2004){Samus}, {Durlevich}, \& {et
  al.}}]{2004yCat.2250....0S}
{Samus}, N.~N., {Durlevich}, O.~V., \& {et al.} 2004, VizieR Online Data
  Catalog, 2250, 0

\bibitem[{{Sharov} {et~al.}(2000){Sharov}, {Alksnis}, {Zharova}, \&
  {Shokin}}]{2000AstL...26..433S}
{Sharov}, A.~S., {Alksnis}, A., {Zharova}, A.~V., \& {Shokin}, Y.~A. 2000,
  Astronomy Letters, 26, 433

\bibitem[{{Shore} {et~al.}(1996){Shore}, {Starrfield}, \&
  {Sonneborn}}]{1996ApJ...463L..21S}
{Shore}, S.~N., {Starrfield}, S., \& {Sonneborn}, G. 1996, \apjl, 463, L21

\bibitem[{{Stanek} \& {Garnavich}(1998)}]{1998ApJ...503L.131S}
{Stanek}, K.~Z. \& {Garnavich}, P.~M. 1998, \apjl, 503, L131

\bibitem[{{Starrfield}(1989)}]{1989clno.conf...39S}
{Starrfield}, S. 1989, in Classical Novae, 39

\bibitem[{{Starrfield} {et~al.}(1974){Starrfield}, {Sparks}, \&
  {Truran}}]{1974ApJS...28..247S}
{Starrfield}, S., {Sparks}, W.~M., \& {Truran}, J.~W. 1974, \apjs, 28, 247

\bibitem[{{Stiele} {et~al.}(2008){Stiele}, {Pietsch}, {Haberl}, \&
  {Freyberg}}]{2008A&A...480..599S}
{Stiele}, H., {Pietsch}, W., {Haberl}, F., \& {Freyberg}, M. 2008, \aap, 480,
  599

\bibitem[{{Supper} {et~al.}(2001){Supper}, {Hasinger}, {Lewin}, {Magnier}, {van
  Paradijs}, {Pietsch}, {Read}, \& {Tr{\"u}mper}}]{2001A&A...373...63S}
{Supper}, R., {Hasinger}, G., {Lewin}, W.~H.~G., {et~al.} 2001, \aap, 373, 63

\bibitem[{{Tuchman} \& {Truran}(1998)}]{1998ApJ...503..381T}
{Tuchman}, Y. \& {Truran}, J.~W. 1998, \apj, 503, 381

\bibitem[{{Warner}(1995)}]{1995cvs..book.....W}
{Warner}, B. 1995, {Cataclysmic variable stars} (Cambridge Astrophysics Series,
  Cambridge, New York: Cambridge University Press, 1995)

\bibitem[{{Williams} {et~al.}(2004){Williams}, {Garcia}, {Kong}, {Primini},
  {King}, {Di Stefano}, \& {Murray}}]{2004ApJ...609..735W}
{Williams}, B.~F., {Garcia}, M.~R., {Kong}, A.~K.~H., {et~al.} 2004, \apj, 609,
  735

\bibitem[{{Williams} {et~al.}(2008){Williams}, {Milne}, {Park}, {Barthelmy},
  {Hartmann}, {Updike}, \& {Hurley}}]{2008AIPC.1000..535W}
{Williams}, G.~G., {Milne}, P.~A., {Park}, H.~S., {et~al.} 2008, in American
  Institute of Physics Conference Series, Vol. 1000, American Institute of
  Physics Conference Series, 535

\bibitem[{{Yaron} {et~al.}(2005){Yaron}, {Prialnik}, {Shara}, \&
  {Kovetz}}]{2005ApJ...623..398Y}
{Yaron}, O., {Prialnik}, D., {Shara}, M.~M., \& {Kovetz}, A. 2005, \apj, 623,
  398

\end{thebibliography}

\end{document}